\documentclass[onecolumn]{mn2e}

\usepackage{amssymb}

\usepackage{amsmath}

\usepackage[dvips]{graphicx}

\usepackage{color}

\title[Primordial Magnetic Fields]{Amplification of Primordial Magnetic Fields by Anisotropic Gravitational Collapse}

\author[Emma J King \& Peter Coles]{Emma J King and Peter Coles\\
School of Physics \& Astronomy, University of Nottingham,
University Park, Nottingham, NG7 2RD, United Kingdom\\ }

\date{}

\begin{document}

\maketitle

\begin{abstract}
If a magnetic field is frozen into a plasma that undergoes spherical
compression then the magnetic field $B$ varies with the plasma density
$\rho$ according to $B \propto \rho^{2/3}$. In the gravitational
collapse of cosmological density perturbations, however,
quasi-spherical evolution is very unlikely. In anisotropic collapses
the magnetic field can be a much steeper function of gas density than
in the isotropic case. We investigate the distribution of
amplifications in realistic gravitational collapses from Gaussian
initial fluctuations using the Zel'dovich approximation. Representing
our results using a relation of the form $B\propto \rho^{\alpha}$, we
show that the median value of $\alpha$ can be much larger than the
$\alpha=2/3$ resulting from spherical collapse, even if there is no
initial correlation between magnetic field and principal collapse
directions. These analytic arguments go some way towards understanding
the results of numerical simulations.
\end{abstract}


\section{Introduction}

One of the most significant gaps in our understanding of cosmological
structure formation is how large-scale magnetic fields arise and how
they relate to the processes involved in galaxy and cluster
formation. The observed galactic fields, with typical strengths of a
few $\mu$G, could be the result of the amplification of a tiny seed
field, perhaps small as $\sim 10^{-20}$G. On the other hand the
adiabatic compression of a somewhat larger primordial field of $\sim
10^{-9}$G could also achieve similar strengths in bound objects
without violating cosmological constraints. For general reviews of
various aspects of cosmological magnetic fields, see Kronberg (1994),
Widrow (2002) and Vallee (2004).

One specific situation where magnetic fields might prove important is
in galaxy clusters. The most convincing line of evidence demonstrating
the existence of magnetic fields in clusters emerges from studies of
Faraday rotation measures (RM). Typical RM values of galaxy clusters
are of the order of a few 100 rad/m$^2$, roughly consistent with field
strengths of a few $\mu$G. These are comparable with galactic magnetic
fields, but well below equipartition with the thermal cluster
gas. Much larger RM values of a few 1000 rad/m$^2$ have been detected
in cooling-flow clusters, indicating possibly substantial magnetic
pressure support of the intra-cluster gas there (e.g. Carilli \&
Taylor 2002). Although the magnetic fields of galaxy clusters are less
ordered than those of spiral galaxies, the presence of coherent
structures is suggested by high resolution Faraday maps (e.g.  Dreher
et al. 1987; Taylor \& Perley 1993; Taylor et al. 2001; Eilek \& Owen
2002). These structures could result from shear-amplification of
originally small-scale magnetic fields, as is suggested by MHD
simulations of galaxy cluster formation (Dolag et al. 1999). It is
clear, however, that the overall power-spectrum of magnetic field
fluctuations in clusters is rather broad ; see Dolag et
al. (2002). The observed magnetic fields could play a significant role
in supporting the intracluster medium through magnetic pressure (Loeb
\& Mao 1994) and there is some evidence that fields may be dynamically
important in some clusters (Dolag et al. 2001; Eilek \& Owen 2002). It
is not known, however, how important these are or how they got there
(Dolag et al. 1999; Goncavles \& Friaca 1999; Dolag \& Schindler
2000; Vlahos et al. 2005).

In this paper we look at the behaviour of a primordial magnetic field
during cosmological evolution. Our aim is to develop an understanding
of how gravitational collapse can, even in the absence of turbulence
or other dynamo action, lead to a significant amplification of
magnetic fields during the the evolution of clusters or other
large-scale cosmic structures. The essence of our argument is very
simple. Assume there is a primordial magnetic field in a region
undergoing compression and that the field is frozen into the plasma in
that region. Since flux is conserved, the magnetic field strength
scales with the cross-sectional area of the collapsing region while
the gas density scales with its volume. The net result for a spherical
compression is that $B \propto \rho^{2/3}$ assuming no back-reaction
of the field on the collapse (which we assume throughout this
paper). One is tempted to infer that the ``average'' amplification due
to collapse of a random set of density perturbations should be equal
to this value. However, it has been known since the pioneering work of
Zel'dovich (1970) and Doroshkevich (1970) that the generic collapse of
cosmological perturbations is very anisotropic; see also Shandarin \&
Zelovich (1989). It is also true that the amplification of the
$B$--field is a non-linear function of the overall collapse
factor. Put these two facts together with the realisation that
averaging does not commute with non-linear operations and it is clear
that the angle-averaged amplification of $B$ need not be close to the
isotropic value. This argument was first presented by Bruni, Maartens
\& Tsagas (2003) who showed that there are possible collapse
geometries that lead to a much larger boost in the $B$--field than one
would naively expect. There is some evidence that this may be observed
in galaxy clusters through a large value of the slope of the $B-n_e$
relation: taking a model of the form $B\propto \rho^\alpha$ it seems
values of $\alpha\sim 0.9$ can be inferred (Dolag et al. 2001). The
particular issue we are interested in tackling is what one can say
about probabilities of these collapses and whether they lead to a
significant effect after averaging over all possible collapse
directions.

Of course there are a number of complicated astrophysical processes
that could boost the magnetic field in a cluster: turbulent dynamo
action and the expulsion of galactic $B$-fields are just two
examples. These are beyond the scope of existing numerical methods, as
well as analytic modelling, but it is nevertheless important to
understand how much of the observed behaviour of cluster magnetic
fields can be explained using simple arguments like those we present
here.

In the next Section we run through some basic definitions and simple
examples of anisotropic collapse. In Section 3 we use an argument
based on the Zel'dovich approximation to calculate the distribution of
$\alpha$ for freely collapsing fluid elements arising from Gaussian
initial fluctuations.

\section{Basics}

\begin{center}
\begin{figure}
\includegraphics{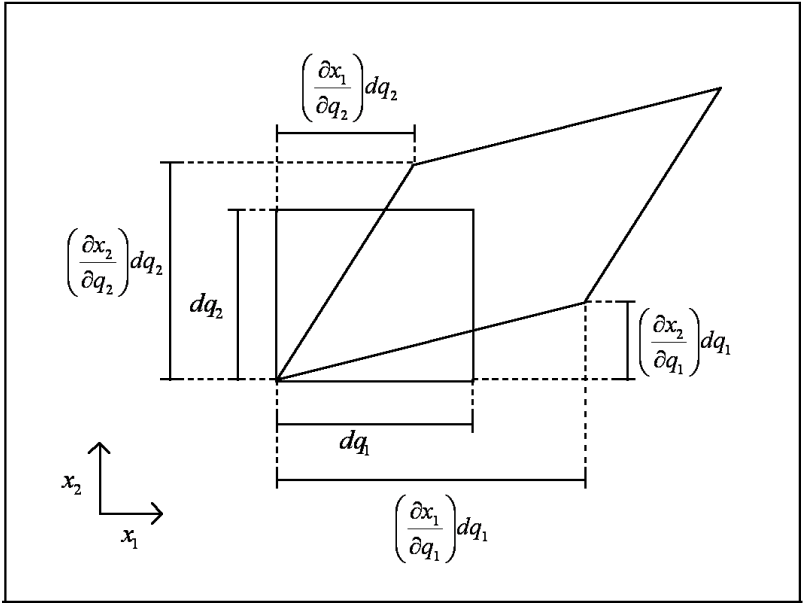}
%\vspace{3.5cm}
\caption{The deformation of a 2-D fluid element of initial side 
lengths $dq_i$, described by the deformation tensor ${\cal D}_{ij} 
\equiv \frac{\partial{x_i}}{\partial{q_j}}$.  The diagonal elements 
of the deformation tensor, ${\cal D}_{11}$ \& ${\cal D}_{22}$
describe the scaling of the fluid element, while 
the off diagonal elements, ${\cal D}_{12}$ \& ${\cal D}_{21}$
describe the skew movement.} 
\label{fig:deformation}
\end{figure}
\end{center}

Consider the effect of a general deformation of a fluid element.
Define a {\em deformation tensor}, ${\cal D}_{ij}$, representing
the deformation of an infinitesimal fluid element of initial side
lengths $dq_i$, such that
\begin{equation}
\label{eq:deformation} {\cal D}_{ij} \equiv
 \frac{\partial{x_i}}{\partial{q_j}}.
\end{equation}
The deformation described by ${\cal D}_{ij}$ is illustrated for the
2-D case in Figure \ref{fig:deformation}.  
The initial co-moving volume of the fluid element, in three
dimensions, is just $dq_1 dq_2 dq_3$, and its final comoving
volume is $\lvert {\cal D}_{ij} \rvert dq_1 dq_2 dq_3$, giving a
ratio of final to initial volume of
\begin{equation}
\label{eq:Vratio} \frac{V^{\rm fin}}{V^{\rm ini}}= \frac{\rho^{\rm
ini}}{\rho^{\rm fin}}= \rvert{\cal D}_{ij}\lvert.
\end{equation}

The strength of a magnetic field is proportional to the density of
its field lines.  Hence, under the assumption that field lines are
frozen into the fluid (as is usually assumed to be the case in
astrophysical situations, due to the large scales involved),
movement of the fluid perpendicular to the field lines affects the
field strength, while movement parallel to the field lines leaves
it unchanged.
\begin{center}
\begin{figure}
\includegraphics{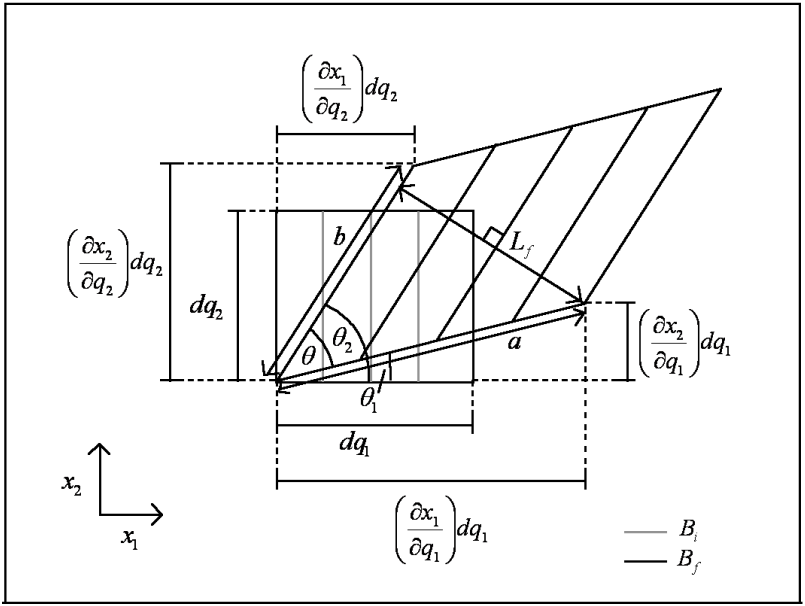}
%\vspace{3.5cm}
\caption{The effect of the deformation of a fluid element on the 
magnetic field lines in 2-D, assuming the field lines are frozen 
into the fluid.  The scaling of the fluid element affects the 
field strength only if the movement is perpendicular to the 
direction of the field lines.  Skew movement affects both the 
field strength and direction.} 
\label{fig:2Ddef}
\end{figure}
\end{center}
In order to see how the deformation tensor relates to the change
in field strength it is easiest first to consider the two
dimensional case, as illustrated in Figure \ref{fig:2Ddef}. We
first take the case where the initial field $\boldsymbol{B}^{\rm
ini} = B^{\rm ini}_2$ only. The strength of the magnetic field is
proportional to the area (or in this 2-D case, length)
perpendicular to the field lines.  Hence in Figure
\ref{fig:2Ddef},
\begin{equation}
B_2^{\rm ini}dq_1 = B^{\rm fin}L_f. \label{eq:Bf}
\end{equation}
We can find $L_f$, the final length perpendicular to the field
lines, by noting that
\begin{equation}
L_f = a \sin(\theta_2 - \theta_1) = a
 \left(\sin(\theta_2)\cos(\theta_1) - \cos(\theta_2)\sin(\theta_1)
\right),
\end{equation}
so from Equation \ref{eq:Bf} the magnitude of the final field,
$B^{\rm fin}$, is given by
\begin{equation}
B^{\rm fin} =\frac{B_2^{\rm ini}
dq_1}{a\left(\sin(\theta_2)\cos(\theta_1)
-\cos(\theta_2)\sin(\theta_1)\right)} =\frac{b}{dq_2}
\frac{B_2^{\rm ini}}{\lvert {\cal D}_{ij} \rvert}
\end{equation}
since
\begin{equation}
\theta_1 = \sin^{-1}\left(\frac{{\cal D}_{12}dq_1}{a}\right) =
\cos^{-1}\left(\frac{{\cal D}_{11}dq_1}{a}\right) \end{equation}
and \begin{equation} \theta_2 = \sin^{-1}\left(\frac{{\cal
D}_{22}dq_2}{b}\right) =\cos^{-1}\left(\frac{{\cal
D}_{12}dq_2}{b}\right), 
\end{equation} 
and in two dimensions
\begin{equation}
{\lvert {\cal D}_{ij} \rvert}=\left({\cal D}_{11}{\cal D}_{22} -
{\cal D}_{12}^2 \right). 
\end{equation} 
Since the final field is
no longer in the $x_2$ direction only we also need to know its
$x_1$ and $x_2$ components, which are given by
\begin{equation}
B_1^{\rm fin} = B^{\rm fin} \cos(\theta_2) = B_2^{\rm
ini}\frac{{\cal D}_{12}}{\lvert {\cal D}_{ij} \rvert}
\end{equation}
and {\begin{equation}
 B_2^{\rm fin} = B^{\rm fin} \sin(\theta_2) =
B_2^{\rm ini}\frac{{\cal D}_{22}}{\lvert {\cal D}_{ij} \rvert}.
\end{equation}
Following the same route for a field which is initially in the
$x_1$ direction only gives similar results, leading to total $x_1$
and $x_2$ components of the final field configuration given by
\begin{equation}
B_1^{\rm fin} = \frac{1}{\lvert {\cal D}_{ij}
\rvert}\left(B_1^{\rm ini} {\cal D}_{11} + B_2^{\rm ini} {\cal
D}_{12} \right), \quad B_2^{\rm fin} = \frac{1}{\lvert {\cal
D}_{ij} \rvert}\left(B_1^{\rm ini} {\cal D}_{12} + B_2^{\rm ini}
{\cal D}_{22} \right).
\end{equation}
Extending this argument to three dimensions gives
\begin{equation}
B_i^{\rm fin}=\frac{1}{\lvert {\cal D}_{ij}
\rvert}\sum_j\left(B_j^{\rm ini}{\cal D}_{ij}\right).
\end{equation}
The total magnitude of a magnetic field is given by
\begin{equation}
\label{eq:Bmag} \lvert \boldsymbol{B} \rvert = \sqrt{(B_1)^2 +
(B_2)^2 + (B_3)^2}
\end{equation}
so in three dimensions the ratio of the magnetic field strengths
before and after deformation is given by
\begin{equation}
\label{eq:Bmagratio} \frac{\lvert \boldsymbol{B}^{\rm fin}
\rvert}{\lvert \boldsymbol{B}^{\rm ini} \rvert} = \frac{1}{\rvert
{\cal D}_{ij} \lvert}\left( \frac{\sum_{i,j} \left(B_i^{\rm
ini}\right)^2 {\cal D}_{ij}^2} {\sum_i \left(B_i^{\rm
ini}\right)^2} \right)^{\frac{1}{2}}.
\end{equation}

In the case of isotropic collapse, ${\cal
D}_{xx} = {\cal D}_{yy} = {\cal D}_{zz} = {\cal D}_{ii}$, and all
the off diagonal elements with $i \neq j$ are zero, giving
\begin{equation}
\frac{\lvert \boldsymbol{B}^{\rm fin} \rvert}{\lvert
\boldsymbol{B}^{\rm ini} \rvert} = \frac{{\cal
D}_{ii}}{\left({\cal D}_{ii}\right)^3} = \left(\frac{\rho^{\rm
fin}}{\rho^{\rm ini}}\right)^{2/3}
\end{equation}
as expected, but the general case, which is not necessarily
isotropic and can also include skew components, is more
complicated than this.

This is all very well, but we need to set up a model that applies
to collapses of fluid elements within a spatially varying random
field of density perturbations. Let us assume, for a toy model,
that each possible direction of collapse is independent and
equally likely. It is then straightforward to work out the average
amplification of the magnetic field which these deformations would
produce. For reasons outlined in the introduction we focus on the
parameter $\alpha$, where \begin{equation} B \propto
\rho^{\alpha}.
\end{equation}
We look first at the case of simple one--dimensional collapse,
without skew components. If we assume that our initial magnetic
field is in the $x_1$ direction only, then for collapse in the
$x_1$ direction we have ${\cal D}_{11}={\cal D}_{11}$. The other
diagonal elements ${\cal D}_{22} = {\cal D}_{33} = 1$, with all
off--diagonal elements ${\cal D}_{ij} (i\neq j) =0$, which gives
\begin{equation}
\frac{\lvert \boldsymbol{B}^{\rm fin} \rvert} {\lvert
\boldsymbol{B}^{\rm ini} \rvert} = \frac{1}{{\cal D}_{11}}
\frac{\sqrt{(B_1^i)^2({\cal D}_{11}^2)}}{\sqrt{(B_1^i)^2}} = 1,
\end{equation}
so that $B  \propto \rho^{0}$ in this case. For collapse in the
$x_2$ or $x_3$ direction only we have, for each case,
\begin{equation}
\frac{\lvert \boldsymbol{B}^{\rm fin} \rvert}{\lvert
\boldsymbol{B}^{\rm ini} \rvert} = \frac{1}{{\cal D}_{ii}}
\end{equation}
In this case we see that  $B \propto 1/V \propto \rho^{1}$.We will
from now on work with averages obtained over all possible
configurations of the deformation by associating a value of
$\alpha$ with each possible geometry and averaging these values.
If these three eventualities discussed above are equally likely
then the average value of $\alpha=2/3$, so that the average effect
is $B \propto \rho^{2/3}$, just as in the isotropic case.

We now look at the three possible skew directions, where the
diagonal elements, ${\cal D}_{ii}=1$, and one of the off--diagonal
elements ${\cal D}_{ij}={\cal D}_{ij}$, with the other
off--diagonal elements $=0$.  When ${\cal D}_{12}$ is the only
non-zero off--diagonal element we have
\begin{equation}
\frac{\lvert \boldsymbol{B}^{\rm fin} \rvert}{\lvert
\boldsymbol{B}^{\rm ini} \rvert} = \frac{1}{1+{\cal D}_{12}}
\frac{\sqrt{(B_1^i)^2+(B_1^i)^2({\cal
D}_{12})^2}}{\sqrt{(B_1^i)^2}} = \frac{1}{(1+{\cal
D}_{12})^{1/2}},
\end{equation}
so that $B \propto \rho^{1/2}$. Similarly, when ${\cal D}_{13}$ is
the only non-zero off diagonal element we also find $B \propto
\rho^{1/2}$, but for the case where ${\cal D}_{23}$ is non-zero,
we find
\begin{equation}
\frac{\lvert \boldsymbol{B}^{\rm fin} \rvert}{\lvert
\boldsymbol{B}^{\rm ini} \rvert} = \frac{1}{(1+{\cal D}_{23})},
\quad B \propto \rho^{1}.
\end{equation}
Again we find that the average over these three possibilities is
$B \propto \rho^{2/3}$, as for the case of isotropic collapse.

However, in reality, each of these possible directions of collapse
is not independent and equally likely.  On the contrary, the
elements of the deformation tensor are correlated, and as a
consequence the amplification of the field is, on average, quite
different from what would be expected from this naive model; see
Bruni et al. (2003). In the next Section we investigate more
realistic collapse geometries.

\section{Realistic Collapse Models}

\subsection{The Zel'dovich Approximation}

The Zel'dovich approximation (Zel'dovich 1970; Doroshkevich 1970;
Shandarin \& Zel'dovich 1989; Bartelmann \& Schneider 1992; Sahni
\& Coles 1995) is an approach to structure formation which assumes
that each particle has some small initial peculiar velocity which
is determined by the initial density fluctuations. Particles then
travel on ballistic trajectories determined by this initial
veloicty, but unaffected by subsequent changes to the
gravitational potential. Hence we write $\boldsymbol{r}(t)$, the
final (Eulerian) position of a particle at time $t$, as
\begin{equation}
\label{eq:zeldovich} \boldsymbol{r} = a(t)\boldsymbol{q} +
a(t)b(t)\boldsymbol{v}(\boldsymbol{q}),
\end{equation}
where $a(t)$ is the scale factor of the universe, $\boldsymbol{q}$
is the initial (Lagrangian) position of the particle, $b(t)$ is
the growing mode, which scales the peculiar velocity,
$\boldsymbol{v}$, with time, and $\boldsymbol{v}$ depends only the
initial position $\boldsymbol{q}$.  We can see that this equation
looks like the background Hubble expansion, $a(t)\boldsymbol{q}$,
plus some perturbation which vanishes as $t \rightarrow 0$.

Changing to co-moving co-ordinates, $\boldsymbol{x} =
\boldsymbol{r}/a(t)$, gives
\begin{equation}
\label{eq:zeldovichcomove} \boldsymbol{x} = \boldsymbol{q} +
b(t)\boldsymbol{v}(\boldsymbol{q}).
\end{equation}
If we assume that the initial velocity is irrotational (a
reasonable assumption since it is generated by the gravitational
effect of overdensities) we can write equation
(\ref{eq:zeldovichcomove}) in terms of a potential, $\phi$, giving
\begin{equation}
\label{eq:zeldovichpot} \boldsymbol{x} = \boldsymbol{q} +
b(t)\nabla \phi(\boldsymbol{q}),
\end{equation}
where $\nabla \phi = \boldsymbol{v}$, and $\nabla$ denotes
differentiation with respect to the initial coordinates,
$\boldsymbol{q}$. This equation looks very much like ballistic
motion,
\begin{equation}
\label{eq:zeldovichtau} \boldsymbol{x} = \boldsymbol{q} + \tau
\nabla \phi(\boldsymbol{q}),
\end{equation}
but with a peculiar time co-ordinate $\tau = b(t)$.

The Zel'dovich deformation tensor, ${\cal D}_{ij}$, describes the
deformation of a fluid element of initial side lengths $dq_i$, and
is given by
\begin{equation}
\label{eq:deformationx} {\cal D}_{ij} =
 \frac{\partial{x_i}}{\partial{q_j}}
= \delta_{ij} + \tau \frac{\partial^2\phi}{\partial q_i
\partial q_j},
\end{equation}
where $\delta_{ij}$ is the Kronecker delta and $i,j=1,2,3$. The
diagonal elements relate to the expansion and contraction of the
fluid element along the principle axes, while the off-diagonal
elements describe the skew distortion. The standard approach
involves diagonalising the deformation tensor (which is
symmetric). Let us denote the eigenvalues of ${\cal D}_{ij}$ by
$\lambda_k$, so that the mass density $\rho({\bf x}, t)$ becomes
\begin{equation}
\rho ({\bf x}, t)= \rho_0(t) \lvert \frac{\partial {\bf
x}}{\partial {\bf q}} \rvert = \rho_0(t) \Pi_{i=1}^{3}
\left[1+b(t)\lambda_i({\bf q})\right]^{-1}, \label{eq:zelpan}
\end{equation}
where $\rho_0(t)$ is the mean cosmological background density.

The Zel'dovich approximation describes the behaviour of evolving
cosmological perturbations until the Jacobian in Equation
(\ref{eq:zelpan}) becomes zero, at which point the matter density
becomes infinite and a {\em caustic} forms, where the mapping
between Eulerian and Lagrangian coordinates is no longer unique.
Although simple, the Zel'dovich approximation gives remarkably
accurate results up to shell crossing (Coles et al. 1993). This,
along with its natural formulation in terms  of the deformation
tensor describing the collapse of each fluid element, makes it an
ideal approach for studying the effect of gravitational collapse
on magnetic field strength.

\subsection{Probabilities}

The joint probability density function (JPDF) for the elements of
the Zel'dovich deformation tensor is given by Doroshkevich (1970),
where a very elegant discussion is also presented about the
probability of various collapse geometries. The most likely
collapse is initially along one direction, resulting in caustics
that are generically two--dimensional (``pancakes''). To proceed
we assume the initial density perturbations are Gaussian, as is
expected from inflationary models (Bardeen et al. 1986). The
initial velocity potential is then also Gaussian, and so are its
derivatives. Therefore the JPDF of the second derivatives of the
velocity potential, on which the elements of the deformation
tensor depends, is a multivariate Gaussian distribution.  If we
write all the possible second derivatives as a vector,
$\varphi_p$, $p=1...6$, with the diagonal elements, which have
$i=j$, as $\varphi_{1}$, $\varphi_{2}$, $\varphi_{3}$ and the
off--diagonal elements with $i \neq j$ as $\varphi_{4}$,
$\varphi_{5}$, $\varphi_{6}$, then their joint probability is
given by
\begin{equation}
\label{eq:multivariategaussian} P(\varphi_1\dots \varphi_6) =
\frac {\lvert C_{pq}^{-1} \rvert^{\frac{1}{2}} \exp
\left[-\frac{1}{2} \sum_{p,q}\varphi_p C_{pq}^{-1} \varphi_q
\right] } { \left(2 \pi \right)^{3}}
\end{equation}
and we need only find the value of the covariance matrix,
$C_{pq}$, to know the probability distribution in full. Note that
all the variates involved have zero mean. The components of
$C_{pq}$ are given by
\begin{equation}
\label{eq:xcorrelation} C_{pq} \equiv \langle \varphi_p \varphi_q
\rangle = \langle \frac{\partial^2 \phi}{\partial q_i
\partial q_j} \frac{\partial^2 \phi}{\partial q_k\partial q_l}\rangle =
\frac{\sigma^2}{15}\left(\delta_{ij}\delta_{kl} +
\delta_{ik}\delta_{jk} + \delta_{il}\delta_{jk} \right),
\end{equation}
where
\begin{equation}
\label{eq:xcorrelationderiv2} \sigma^2 = \frac{1}{2 \pi^2} \int
P(k) k^2 dk,
\end{equation}
if $P(k)$ is the power spectrum. So we now know the multivariate
Gaussian distribution which gives the JPDF of the elements of the
deformation tensor, provided we know $P(k)$, the power spectrum of
the initial density perturbations. For simplicity we take the case
where the magnetic field is entirely independent of the
deformation tensor (i.e. we are not appealing to the field to
create the initial deformation in any way, so there is no
correlation between the field and the deformation), and ignore any
back reaction of the field on the fluid.  This gives us the
freedom to choose to have the initial magnetic field along one
axis, eg $\boldsymbol{B}^{\rm ini} = B_1^{\rm ini}$ only.  This
simplifies the equation for the field strength amplification
(Equation \ref{eq:Bmagratio}) considerably, giving
\begin{equation}
f_B\equiv\frac{\lvert \boldsymbol{B}^{\rm fin} \rvert}{\lvert
\boldsymbol{B}^{\rm ini} \rvert}
 = \frac{1}{\rvert {\cal D}_{ij} \lvert}
\left({\cal D}_{11}^2 + {\cal D}_{12}^2 + {\cal
D}_{13}^2\right)^{1/2}.
\end{equation}
Since the marginal distribution $P(x) = \int P(x|y)P(y) dy$, the
probability distribution of the magnetic field amplification,
given the deformation tensor, is given by
\begin{equation}
\label{eq:disB} P\left(f_B=\frac{\lvert \boldsymbol{B}^{\rm fin}
\rvert}{\lvert \boldsymbol{B}^{\rm ini} \rvert}\right) =
\int_{-\infty}^{\infty} \delta_D\left(\frac{1}{\rvert {\cal
D}_{ij} \lvert}\left({\cal D}_{11}^2 + {\cal D}_{12}^2 + {\cal
D}_{13}^2\right)^{1/2} - \frac{\lvert \boldsymbol{B}^{\rm fin}
\rvert}{\lvert \boldsymbol{B}^{\rm ini} \rvert}\right) \frac
{\lvert C_{pq}^{-1} \rvert ^{\frac{1}{2}} \exp \left[-\frac{1}{2}
\varphi_p^TC_{pq}^{-1} \varphi_q \right]} { \left(2 \pi
\right)^{3}} d\varphi_{1 \dots 6}
\end{equation}
where
\begin{equation}
\lvert C_{pq}^{-1} \rvert  =
\frac{1}{20}\left(\frac{15}{\sigma^2}\right)^6,
\end{equation}
and $\delta_D(x)$ is the Dirac delta-function. Similarly, the the
density compression factor is
\begin{equation}
f_\rho\equiv \frac{\lvert \rho^{\rm fin} \rvert}{\lvert \rho^{\rm
ini} \rvert}
\end{equation}
and its probability distribution is
\begin{equation}
\label{eq:disrho} P\left(f_\rho=\frac{\lvert \rho^{\rm fin}
\rvert}{\lvert \rho^{\rm ini} \rvert}\right) =
\int_{-\infty}^{\infty}d\varphi_{1 \dots 6}
\delta\left(\frac{1}{\rvert {\cal D}_{ij} \lvert} - \frac{\lvert
\rho^{\rm fin} \rvert}{\lvert \rho^{\rm ini} \rvert}\right) \frac
{\lvert C_{pq}^{-1} \rvert^{\frac{1}{2}} \exp \left[-\frac{1}{2}
\varphi_p^TC_{pq}^{-1} \varphi_q \right]} { \left(2 \pi
\right)^{3}} d\varphi_{1 \dots 6}.
\end{equation}
The two distributions (\ref{eq:disB}) and (\ref{eq:disrho}) are
both skewed showing that there are amplifications and compressions
much larger than average. This skewness also increases with time,
as the perturbations become more non-linear. This suggests that
there is a significant probability of large amplifications of $B$
relative to $\rho$, as suggested by Bruni et al. (2003). However
these two distributions are not sufficient to calculate the
distribution of our chosen parameter $\alpha$, which we define by
\begin{equation} \alpha \equiv \frac{\log (f_B)}{\log (f_\rho)},
\end{equation}
which requires the full joint distribution $P(f_B,f_{\rho})$.
Unfortunately the integrals involved in the construction of this
joint distribution are resistant to analytical methods. In order
to investigate the statistical properties of $\alpha$ (where $B
\propto \rho^{\alpha}$),  we therefore resorted to numerical
techniques.

\subsection{Calculations}

Although the multivariate distributions described above are
difficult to handle analytically, it is fortunately quite easy to
obtain the distribution of $\alpha$ using a Monte Carlo technique.
First we need to generate deformation tensors from the
distribution (\ref{eq:multivariategaussian}). One can generate
correlated Gaussian variates with a given covariance matrix by
simply reversing the process by which one can diagonalise the
covariance matrix to produce independent variates from correlated
ones. In other words, we start with a set of six independent and
identically distributed Gaussian variables, and then construct a
linear combination such that they produce the correct covariance
matrix. This is easily done using a Cholesky decomposition. We use
this technique to generate a large library of deformation tensors
that constitute a fair sample from the required distribution
(\ref{eq:multivariategaussian}). We then used the machinery
described above to calculate the compression factors $f_B$ and
$f_{\rho}$ from this sample. This in turn generates a value of
$\alpha$ for each fluid element from which the overall
distribution can be formed. For simplicity we have assumed, as
before, that there is no correlation between the magnetic field
and the deformation, and have ignored any back reaction of the
magnetic field on the fluid.

There are a number of subtleties that must be confronted before
dealing with this distribution. The first is that not all
deformation tensors compatible with our parent distribution lead
to collapsing fluid elements. Since we wish to look at collapsed
regions only, we consequently restrict our library of deformations
describing a net compression in comoving coordinates. Secondly, since
our variates are Gaussian the eigenvalues $\lambda_i$ are also
Gaussian,  Equation (\ref{eq:zelpan}) always assigns a non-zero value to
regions where shell--crossing has occurred. This generates a tail
in the distribution that is unphysical. For this reason we
illustrate our results by focussing on the field strength
amplification which would be generated by a freely collapsing
fluid element whose overdensity, $\delta$, has reached a value
\begin{equation}
\delta=\frac{\rho-\left<\rho\right>}{\left<\rho\right>}=1.
\end{equation}
Each fluid element in our library was allowed to collapse to this
state, at which point the magnetic field compression factor, $f_B$,
which depends strongly upon the exact configuration of the collapse,
was determined using methods given in Section 2.  This was then
compared to the change in density to find a value of $\alpha$ for each
element. These values were binned to give final distributions of $f_B$
and $\alpha$ which can be seen in Figures \ref{fig:P(B)} and
\ref{fig:P(a)}, respectively.  Notice that the distributions of both B
and $\alpha$ have a very long tail of values. This effect is
understandable in terms of a simple model: if a fluid element
collapses in directions orthogonal to the magnetic field, the field
obviously undergoes a large boost. However, it is possible for the
fluid element to be expanding along the third direction (along the
field). This means that $B$ can increase significantly while $\rho$
hardly changes at all. This can lead to unbounded values of
$\alpha$. These are unlikely, but not impossible so they have a finite
probability. The {\em mean} value of $\alpha$ is consequently formally
undefined.
\begin{figure}
\includegraphics{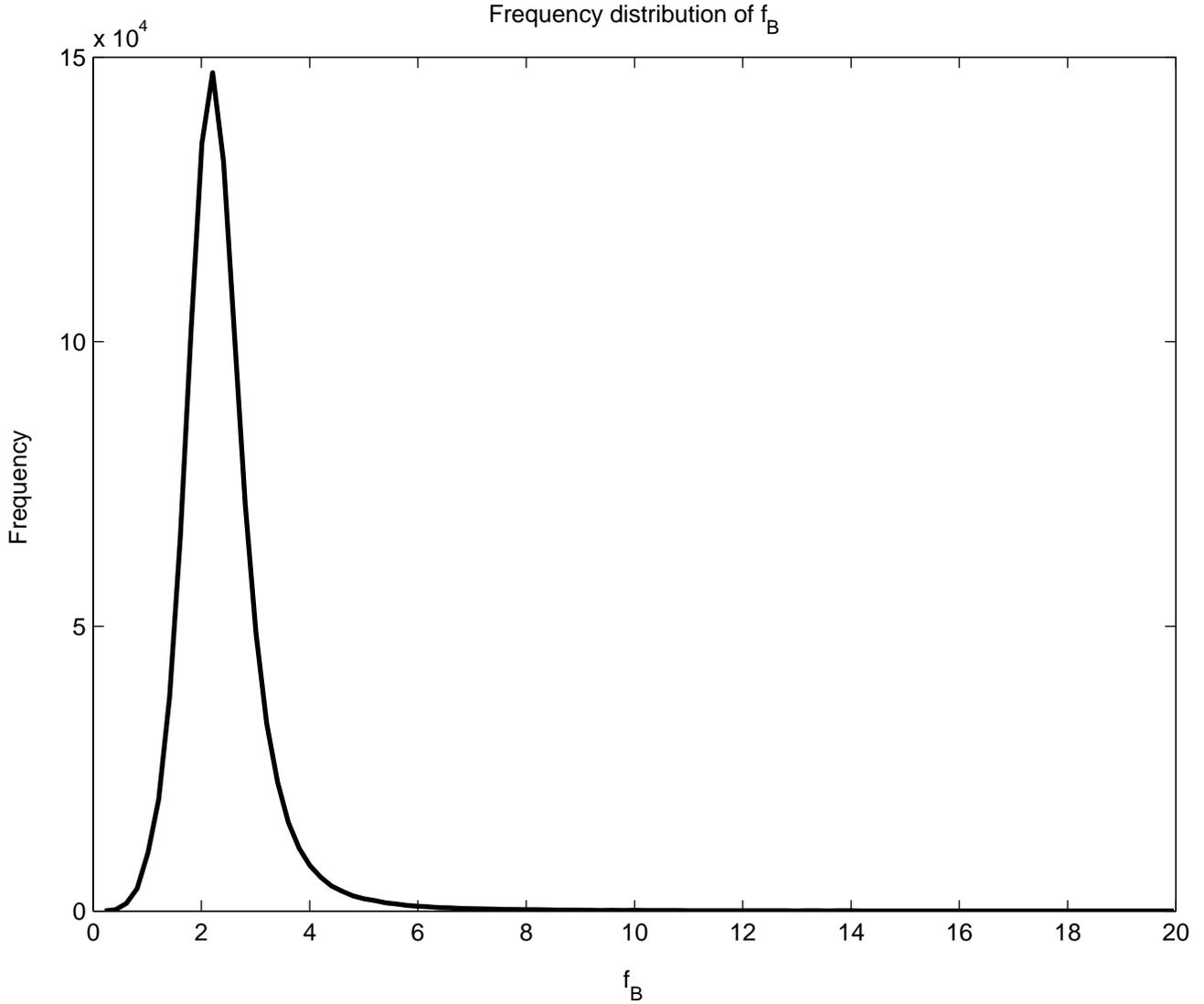}
%\vspace{3.5cm}
\caption{Frequency distribution of $f_B$ when $\delta=1$.  The long tail of
the distribution (which continues to the right of what is shown
above) is due to chance alignments of the direction of collapse with
the magnetic field lines which have a low but non-zero probability.}
\label{fig:P(B)}
\end{figure}
\begin{figure}
\includegraphics{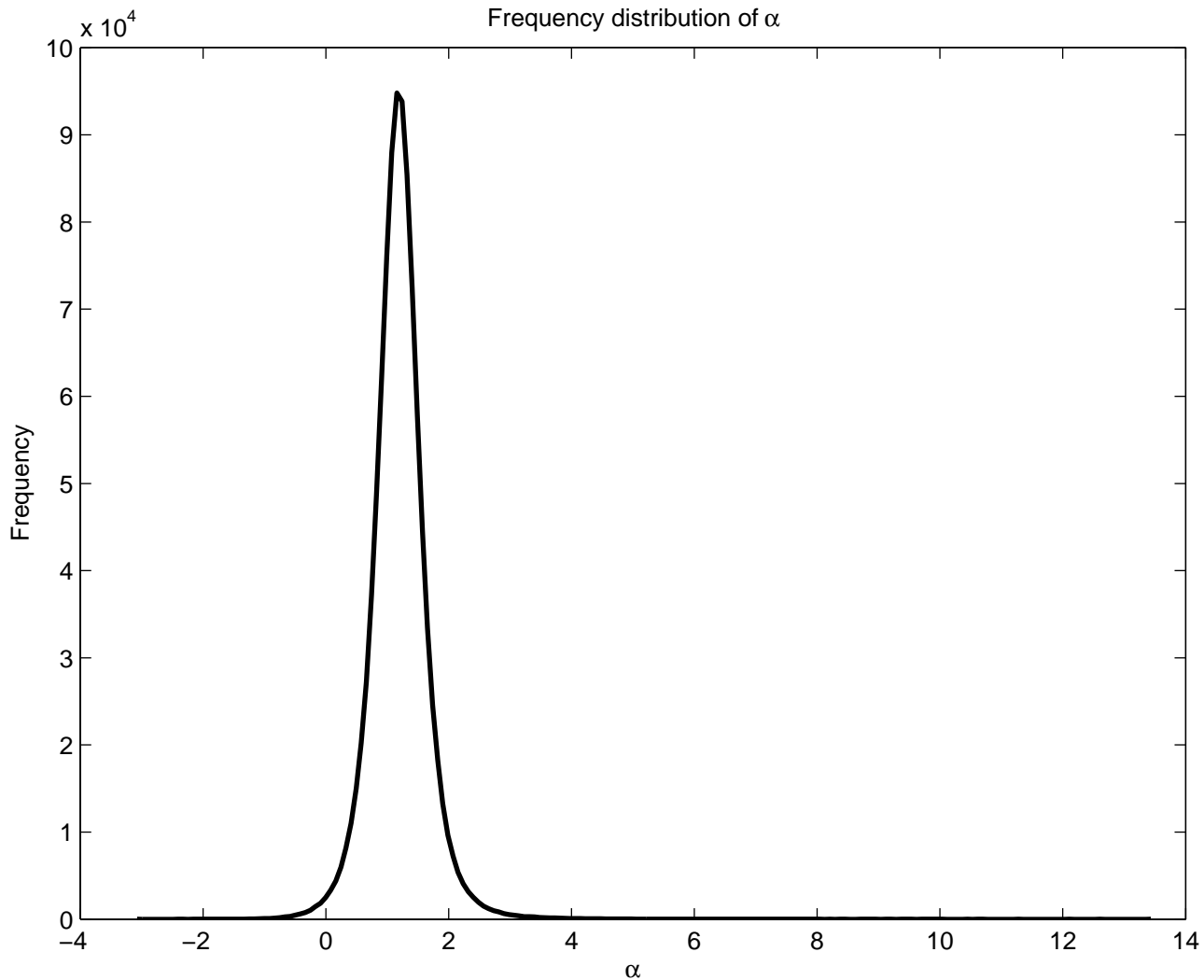}
%\vspace{3.5cm}
\caption{Frequency distribution of $\alpha$ when $\delta=1$.  The mean value of
$\alpha$ is undefined due to the long tail in the distribution, which
is caused by chance alignments of the direction of collapse with the
magnetic field lines.  The median value of $\alpha$ is $1.19$.}
\label{fig:P(a)}
\end{figure}

Intriguingly, the {\em median} value of $\alpha$ once all the
fluid elements had collapsed until $\delta = 1$, is found at a
value of $\alpha=1.19$, considerably higher than the expected
isotropic value of $\alpha=0.67$. One has to be careful
interpreting this result as it does not correspond to a
calculation of $\alpha$ at the same epoch. Some fluid elements
will reach $\delta=1$ at different times to others. Since the
background $B\propto 1/a^2$ and the background $\rho \propto
1/a^3$ the initial ratio effectively depends on the time taken to
collapse through $a(t)$ or, equivalently as $\rho_0^{1/3}(t)$. To
compare these results at the same epoch one should therefore
subtract $1/3$ from the result above giving $\alpha \simeq 0.86$.

While most deformations cause a field amplification at or near the
median value (approximately 87\% are within 50\% of the median),
there are some outliers caused by unusual alignments of the
magnetic field compared to the direction of greatest collapse as
discussed above. High values of $\alpha$ are caused by
configurations where the collapse of the deformation tensor is
such that the magnetic field is amplified greatly for only a small
change in the density. This occurs when the magnetic field is
perpendicular to the axis of fastest collapse, and parallel to an
axis along which the collapse is slow, or expansion is occurring.
On the other hand, if this happens the magnetic field may reach
such a large level that it begins to exert a back-reaction on the
collapse. Our treatment can not handle this eventuality
accurately.

Similarly, extremely low values of $\alpha$, some of which are
negative, occur when the field is perpendicular to an axis which
is collapsing slowly, or even expanding, and parallel to a quickly
collapsing axis, causing a large increase in density for very
small increase, or even a decrease, in magnetic field strength.
Such alignments are unusual, but do occasionally occur, and cause
the long tails in the distribution of $\alpha$ seen in Figure
\ref{fig:P(a)}.

\section{Discussion}

We have investigated the probable magnetic field amplification by
anisotropic gravitational collapse under a number of restrictive
assumptions. These are: \begin{enumerate} \item the initial fluid
perturbations are Gaussian; \item the initial magnetic field is
uncorrelated with the perturbations; \item collapse is described
by the (quasi--linear) Zel'dovich approximation;  \item
shell-crossing does not occur; \item there is no back-reaction of
the magnetic field on the fluid motion.
\end{enumerate}

We find the median value of $\alpha$ to be significantly higher
than the isotropic value of $\alpha=2/3=0.67$, but the precise
value depends on exactly how one performs averages. We have taken
two definitions (which differ by $1/3$ in $\alpha$) but other ways
of averaging are possible. Indeed, for comparison with a single
cosmological object one would probably have to average over
correlated fluid elements, and average by mass rather than by
direction. Obviously if one averages over coherent structures
and/or if there are correlations between magnetic field and the
fluid perturbations, such as would be the case if the field were
generated by fluid activity, then the results would be very
different.

In the numerical simulations presented by Dolag et al. (2001) it
seems values of $\alpha \simeq 0.9$ can be reached. We cannot
argue that our simplified treatment can be compared directly with
these more accurate computations, but it is reassuring that taking
proper account of collapse geometries leads to plausible values of
$\alpha$ that are not far from those reached in fully MHD
approaches. This suggests at any rate that it may be possible to
get some insight into the behaviour of cluster magnetic fields
using simple arguments like those we have presented here,  without
needing to appeal to more complex mechanisms such as sheer flows
and dynamos.

\end{document}